# Comparison of Analytical and Numerical Models for Point to Ring Electro-Hydrodynamic Flow

Yifei Guan[1], Ravi Sankar Vaddi[1], Alberto Aliseda[1], and Igor Novosselov [1,2]
[1]Department of Mechanical Engineering, University of Washington, Seattle, U.S.A. 98195
[2]Institute for Nano-Engineered Systems, University of Washington, Seattle, U.S.A. 98195

*Abstract*—An electrohydrodynamic (EHD) flow in a point-to-ring corona configuration is investigated experimentally, analytically and via a multiphysics numerical model. The interaction between the accelerated ions and the neutral gas molecules is modeled as an external body force in the Navier-Stokes equation (NSE). The gas flow characteristics are solved from conservation principles with spectral methods. The analytical and numerical simulation results are compared against experimental measurements of the cathode voltage, ion concentration, and velocity profiles. A nondimensional parameter, *X*, is formulated as the ratio of the local electric force to the inertial term in the NSE. In the region of $X \geq 1$, the electric force dominates the flow dynamics, while in the $X \ll 1$ region, the balance of viscous and inertial terms yields traditional pipe flow characteristics.

## I. Introduction

Electrohydrodynamic (EHD) flow, also referred to as ionic wind in the literature, has many practical applications, such as convective cooling [1-5], electrostatic precipitators (ESP) [6-9], plasma-assisted combustion [10], airflow control [11], turbulent boundary layer actuators [12], surface particle trapping [13] and collection [14], and electroconvection [15, 16]. A high-voltage corona discharge generates streams of ions between the two electrodes, and the high-velocity ions transfer their kinetic energy to the neutral air molecules outside the corona through collisions, accelerating the gas in the direction of the ion drift. The ions' interaction with the neutral molecules can be modeled as an external force term (Lorentz force) in the Navier-Stokes equations [17, 18]. Insights into the multiphysics nature of the EHD flow are important for understanding this phenomenon. To correctly predict the flow established by this force, the following elements need to be considered: (1) the electric field resulting from the potential difference between the corona and ground electrodes, as well as its modifications due to the space charge in the high ion concentration in the region; (2) the ion motion in the resulting electric field; (3) the interaction between the ion drift and the neutral gas in the flow acceleration region; (4) the viscous and turbulent stresses; and (5) the effects of developing complex flow patterns as a result of the accelerating flow and device geometry.

An analytical model for corona-driven EHD flow and validate the model against the experimental measurements in air. The model addresses the flow acceleration behavior resulting from ion collisions with neutral air molecules in the ion drift region. During the development, we first obtain the relationship between the electrical properties of the EHD flow, such as corona voltage $\varphi$, electric field **E**, and charge density $\rho_c$ for spherical coordinates. Then, the EHD velocity profiles are solved numerically using a Chebyshev spectral method.

To gain insight into the developing EHD flow, a numerical approach is used in a computational fluid dynamics (CFD) simulation that solves for the coupled flow and electric fields in the presence of corona discharge. the scientific literature evaluates several corona configurations [19-21]. Numerical modeling has been applied to the design and analysis of electrostatic precipitators (ESP) [22-25] and heat transfer enhancement [1-5]. Previous EHD flow models use an iterative approach to (1) calculate the electric field and electric force under Kaptzov's hypothesis [26] or Peek's law [27], and (2) set a constant space charge on the anode so that the solution matches the cathode current from the experimental data. This method requires multiple iterations and is therefore inefficient. In contrast, our modeling approach solves for charge density by introducing a volumetric charge flux derived from the anode current directly. The charge flux is imposed on a "numerical ionization region" determined by the electric field and the thresholds for the onset of ionization. The ionization (charge flux) and spatial charge density are two-way coupled to the NSE solver, avoiding the iterative procedure for solving the electric field. The electric force acts on a volume of fluid, inducing the EHD flow; this ion - bulk flow coupling is similar to previous work [1-5, 19-25].

---
[1] ivn@uw.edu

In this manuscript, we compare analytical and numerical models for corona-driven EHD flow. The models are validated against the experimental measurements. The conceptual representation of the EHD system includes (i) gas ionization region, (ii) flow acceleration region where unipolar ion motion in the gas medium acts as a body force accelerating the flow, and (iii) momentum conservation region dominated by the inertial and viscous terms of the NSE. These regions do not necessarily have clear boundaries; but rather, they are characterized based on the flow non-dimensional parameters dominant in each of them. We demonstrate a numerical approach for EHD flow in a finite volume solver for axisymmetric point-to-ring corona configurations. CFD simulations are used to resolve the spatiotemporal characteristics of the flow, electric fields, and charge density. The nondimensional analysis provides insight into the dominant terms in the different EHD flow regions. The electric to kinetic energy transfer efficiency is evaluated for both the model and the experiments.

## II. METHODOLOGY

Experimental measurements for point-to-ring corona discharge are for model development and validation. In the experiments, the two main relationships sought after are voltage-current $(\varphi - I)$, obtained from the anode and cathode, and voltage-velocity $(\varphi - u)$, based on the velocity measurements at the exit of the device. The EHD flow is generated between a charged needle and a grounded ring. The anode needle is a 0.5 mm thick tungsten wire with a radius of curvature at the tip of 1 μm (measured by optical microscopy). It was previously shown that the sharpness of the needle affects the corona discharge at low voltages [28]. To ensure that the needle does not degrade over time, the tip of the needle was inspected routinely. The metal ring is a 1.58 mm thick solid solder with an inner radius of 10 mm. The enclosure is 3D printed with the polylactic acid filament. The air gap between the needle and the ring ($L$) was varied from 3 mm to 7 mm using 3D printed spacers. High voltage positive DC power supply (Bertan 205B-20R) is used to create the electric potential between the needle and the ring.

The measurements were collected for positive corona mode, the ambient temperature in a range of 22-25⁰C, relative humidity of 23-25%, and ambient pressure. For each anode-cathode distance, the voltage was increased from 4 kV (when the outlet velocity is measurable) to ~10 kV (when the arc discharge occurs). Constant current hot-wire anemometry was used to measure the flow velocity profile. A TSI 1213-20 probe connected to the anemometer (AA-1005) was positioned at the outlet of the device. The anemometer probe is calibrated in the range of 0.2 m/s to 8 m/s according to standard procedures [29]. The experimental setup was mounted on an optical table with the anemometer probe attached to the three-dimensional optical stage to provided space-resolved measurements. All components and the probe are grounded. The data from the anemometer is sampled at a frequency of 10 kHz using a data acquisition system (myRIO-1900) for a sampling time of 30 seconds. Each experimental condition is tested at least five times to obtain independent statistical samples. For each distance L, the applied high voltage increases from 4 kV, where the outlet velocity is measurable to around 10 kV or at the onset of the arc discharge.

**Analytical Model**

The simplified ions transport equation can be written as [30]

$$\frac{\mu_b}{\varepsilon} \rho_c^2 - \mu_b \nabla \varphi \nabla \rho_c = 0. \qquad (1)$$

**Table I** shows the solutions for this nonlinear differential equation in spherical coordinates systems; $r$ is the distance from the anode [mm].

*Table I. Solutions for ion transport equation in one dimensional spherical coordinate*

| Variables | Spherical coordinates |
|---|---|
| $\rho_c$ | $\rho_c = \rho_0 r^{-3/2}$ |
| $E = |\mathbf{E}|$ | $E = \dfrac{2\rho_0}{3\varepsilon} r^{-1/2}$ |
| $\varphi$ | $\varphi_0 - \dfrac{4\rho_0}{3\varepsilon} r^{1/2}$ |
| $\rho_0$ | $\dfrac{C}{m^3}\left(mm^{3/2}\right)$ |

The equations for solving maximum axial velocity and velocity profile are given as

$$u_{\max} = \left[ \frac{3\varepsilon}{4\rho r_{cr}} \int_b^{r_{cr}} r^{-2} dr \right]^{1/2} (\varphi_0 - \varphi_{cr}). \qquad (2)$$

$$r \frac{\partial^2 u_a}{\partial r^2} + \frac{\partial u_a}{\partial r} = -\frac{J_0 L^3}{\mu \mu_b} \left[ \frac{r}{(r^2 + L^2)^{3/2}} \right]. \qquad (3)$$

where $u_{\max}$ is the maximum velocity, $u_a$ is the axial velocity, $r$ is the radial dimension, $L$ is the distance from the point to the center of the cathode ring, $J_0$ is the current density, $r_{cr}$ is the critical length, and $\varphi_{cr}$ is the corresponding critical voltage[30]. The Eq. 2-3 are solved to provide maximum velocity and velocity profiles as shown in **FIG. 1** and **FIG. 2**.

**Numerical Model**

In the CFD simulation, the effect of the ion motion interaction on the bulk flow is modeled by adding a body force (electric force) $\mathbf{F_e} = -\rho_e \nabla \varphi$ to the momentum equations. The governing equations used to model the flow are:

$$\nabla \cdot \mathbf{u} = 0, \qquad (4)$$

$$\rho \frac{D\mathbf{u}}{Dt} = -\nabla P + \mu \nabla^2 \mathbf{u} - \rho_e \nabla \varphi, \qquad (5)$$

where $\rho$, the air density ($1.205$ kg/m$^3$), and $\mu$, the air dynamic viscosity [$1.846 \times 10^{-5}$ kg/(ms)], are constant for incompressible isothermal flow, $\mathbf{u} = (u_{axial}, u_{radial})$ is the velocity vector in the two-dimensional axisymmetric model, and $P$ is the static pressure. The ion transport is described by the charge density equation:

$$\frac{\partial \rho_e}{\partial t} + \nabla \cdot \left[ (\mathbf{u} - \mu_b \nabla \varphi) \rho_e - D_e \nabla \rho_e \right] = S_e, \qquad (6)$$

Note that the $-\nabla \varphi = \mathbf{E}$. The electric potential is solved using Gauss's law:

$$\nabla^2 \varphi = -\frac{\rho_e}{\varepsilon_0}, \qquad (7)$$

where $\mu_b$ is the ion mobility, which is approximated as a constant [$2.0 \times 10^{-4}$ m$^2$/(Vs)] at standard pressure and temperature [31, 32], and $\varepsilon_0$ [~$8.854 \times 10^{-12}$ C/(Vm)] is the electric permittivity of free space. Since the relative permittivity of air is close to unity (~1.00059) [33], vacuum permittivity is used in all simulations. $D_e$ is the ion diffusivity described by the electrical mobility equation (Einstein's relation):

$$D_e = \frac{\mu_b k_B T}{q}, \qquad (8)$$

where $k_B$ is Boltzmann's constant (~$1.381 \times 10^{-23}$ J/K), $T$ is the absolute temperature, and $q$ is the electrical charge of an ion, which is equal to the elementary charge ($1.602 \times 10^{-19}$ C).

Ionization is modeled by a volumetric charge flux applied to the fluid within a numerical ionization zone. Instead of defining a surface within the computational domain to model the ionization zone boundary, the volume of the ionization zone is calculated based on the electric field strength computed in the simulation. In Eq. (3), $S_e$ is the volumetric flux term for charge density with units of C/(m$^3$s):

$$S_e = \begin{cases} I / \Psi, \text{ for } |\mathbf{E}| \in [E_0, E_1] \text{ \& } x_{tip} - x < 1 \text{ mm} \\ 0, \text{ otherwise} \end{cases}, \qquad (9)$$

where $\Psi$ is the volume of the numerical ionization zone, calculated in the simulation, it satisfies $|\mathbf{E}| \in [E_0, E_1] \text{ \& } x_{tip} - x < 1$ mm; $I$ is the anode current, measured experimentally and used as a boundary condition in the numerical simulation. The condition $x_{tip} - x$ term limits ion production along the needle. It is based on the experimental electrode setup – the tip extends 1 mm from the needle holder. $E_0 = 2.8 \times 10^6$ V/m is the critical field strength below which the number of ion recombination events is greater than the production per drift length for air[34]. $E_1 = 3.23 \times 10^6$ V/m is the breakdown electric field strength for air [1, 9]. Since the charge density is balanced in the ionization region, the anode current equals the charge density flux at the ionization boundary.

**System Parameterization**

The non-dimensional governing equation is given by

$$\text{St} \frac{\partial \mathbf{u}^*}{\partial t^*} + (\mathbf{u}^* \cdot \nabla^*) \mathbf{u}^* = -\nabla^* P^* + \frac{1}{\text{Fr}^2} \mathbf{g}^* + \frac{1}{\text{Re}} \nabla^{*2} \mathbf{u}^* - \left[ \frac{\rho_e \varphi}{\rho \mathbf{u}^2} \right] \rho_e^* \nabla^* \varphi^* \qquad (10)$$

where St is the Strouhal number, Fr is the Froude number, the asterisk superscript denotes nondimensional variables [35]. A proposed nondimensional parameter, $X = \rho_e \varphi / \rho \mathbf{u}^2$, is defined as the ratio of electrostatic to inertial terms. In global terms, the parameter $X$ is related to the electro-inertial number $N_{EI} = \varepsilon |\vec{E}|^2 / \rho \mathbf{u}^2$ [36], described in the literature as $\text{Md}/\text{Re}^2$, where $|\mathbf{E}|$ is the magnitude of the electric field vector, and Md is the Masuda number [37]. The parallels come from the electric description based on Gauss's law. Gauss's law can be written in a nondimensional form as $\varepsilon |\mathbf{E}|^2 = \rho_e \varphi$ [34].

III. RESULTS AND DISCUSSION

As shown in **FIG. 1**, the numerical model agrees within 5% error with the experimental data and the analytical predictions. The analytical predictions are generally higher than the numerical results since it does not account for viscous drag or the nonunidirectional flow. For a given point-to-ring distance, the maximum outlet velocity increases linearly with corona voltage.

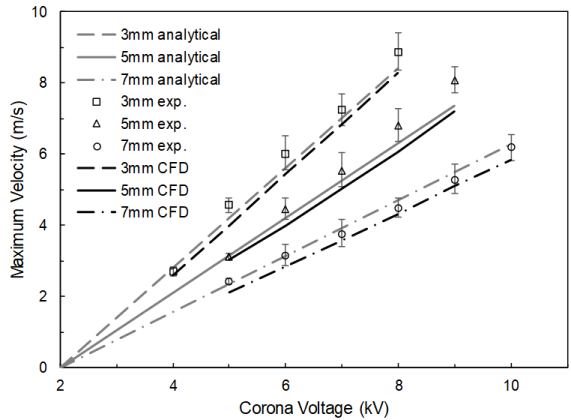

**FIG. 1.** *Maximum velocity as a function of corona voltage and electrode geometry for the experimental, analytical, and CFD results.*

As shown in **FIG. 2**, both the analytical solution and the experimental results show velocity profiles with a very distinct peak at the axis of the coaxial configuration, consistent with the localized electro-hydrodynamic force at the tip of the needle electrode. The velocity profiles then decay quickly over a short radial distance (of the order of the ionization zone width, $r_{cr}/4$) with asymptotic decay towards the edge of the domain, consistent with entrainment in a confined flow environment. The comparison between the analytical solution and the data is excellent at the centerline; the velocity decays approaching the wall is not captured well by the model due to the fully developed assumption implicit in the model. The balance of the viscous stress term by the EHD forcing at the center of the analytical simplification means that the model assumes the convective term to be negligible. This is not valid in the region where the pipe flow, upstream of the corona discharge, must adapt to the new conditions presented by the EHD forcing near the axis. Additionally, the one-dimensional flow assumption cannot describe the formation of more complex flow patterns in the EHD device, which can form due to adverse pressure and electric field gradients. Here, the EHD force is applied only in the axial direction where it captures well the flow acceleration region near the center line but neglects the effect of the three-dimensional nature of the electric field downstream of the cathode.

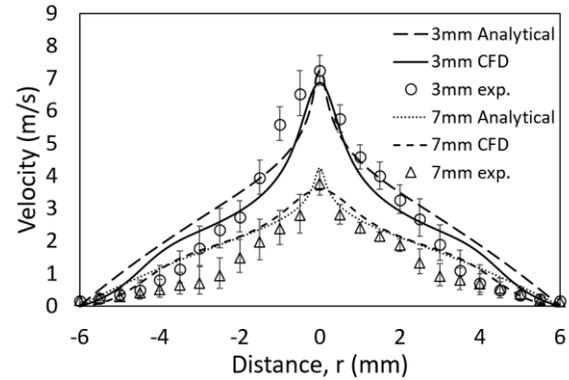

**FIG. 2.** *Comparison between the analytical, numerical and the experimental velocity profiles at the outlet of the EHD generator; corona distances are 3mm and 7mm, fixed corona voltage at 7kv*

The velocity profiles then decay rapidly with radial distance. Numerical and experimental results agree within 5% error at the centerline, but the model is less accurate at the edges of the domain. The discrepancy in this region may be due to flow instability in the shear flow region that modifies the radial location of the inflection points in the velocity profile.

FIG. 3 shows electric field lines colored by the values of $X$, indicating the regions where the electric force is higher than the inertial force. The EHD-dominated flow (red) is located between the corona and ground electrode where both the ion concentration and the electric field strength are high.

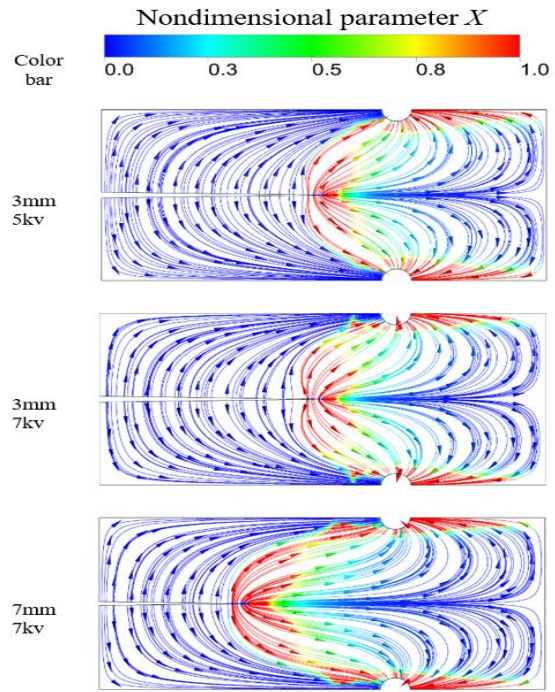

FIG. 3. *Electric field lines colored by the nondimensional parameter $X$. The red zone, $X \geq 1$, indicates the regions of EHD-dominated flow. The color map is limited to $X = 1$; the value $X$ is as high as 400 in the regions near the ionization zone and the low-velocity region near the wall.*

## V. Conclusion

The velocity profile predicted by the analytical model has good agreement near the centerline region of the EHD generator. The analytical model over-predicts the gas velocity near the edge of the domain. The limitations of the model are likely the results of the simplified assumptions in the flow and electric field: (i) the application of the EHD force in the axial direction neglects the effect of the three dimensional nature of the electric field that can result in the formation of complex flow patterns; and (ii) the EHD flow generation model needs to be divided into an pure ion acceleration region model and an inertial flow section where the flow develops under the triple balance between EHD forcing, convective flow acceleration, and viscous shear stresses to capture the transition between the wall-bounded pipe flow and the EHD-driven centerline. The numerical model takes into account the effect of viscous stresses near the walls, as well as the balance between inertia and electric forces; it captures the experimental velocity profile better than the analytical model that provides accurate predictions near the centerline.

## Acknowledgment

This research was supported by the DHS Science and Technology Directorate and UK Home Office, contract no. HSHQDC-15-531 and by the National Institutes of Health, (grant numbers: NIBIB U01 EB021923, NIBIB R42ES026532 subcontract to UW)